\documentclass[prb,aps,floatfix,twocolumn,preprintnumbers,amsmath,amssymb]{revtex4}
\usepackage{graphicx}
\usepackage{dcolumn}
\usepackage{bm}
\usepackage{longtable}

\begin{document}

\title{Rabi-type oscillations in a classical Josephson junction}

\author{Niels Gr{\o}nbech-Jensen}

\affiliation{Department of Applied Science, University of California, Davis,
California 95616}
\author{Matteo Cirillo}
\affiliation{Dipartmento di Fisica and INFM, Universit\`a di Roma
''Tor Vergata'', I-00173 Roma, Italy}

\date{\today}

\begin{abstract}
We study analytically and numerically the phase-modulation properties of a
biased classical Josephson tunnel junction in the zero-voltage state and
phase-locked to an external ac field. We show that the phase-locked state is
being modulated in the transients, or in response to perturbations, and the
modulation frequency is calculated as a function of relevant system
parameters, such as microwave field amplitude. The numerical analysis is
parameterized similarly to recent experimental results in which a
combination of pulsed ac signals and relative switching from the
zero-voltage state are used to probe the internal excitations of the
junctions. Our analysis demonstrates that the modulation of a phase-locked
state in an entirely classical Josephson junction produces oscillations
analogous to quantum mechanical Rabi oscillations, expected to be observed
under the same conditions.
\end{abstract}
\pacs{74.50.+r, 03.67.Lx, 85.25.Cp}

\maketitle

\setcounter{equation}{0}
The symmetric oscillations of a quantum system between its ground and
excited states have recently attracted much interest within the field of
quantum coherence and quantum computation; see, e.g., \cite{silve}.
Existence of such oscillations was originally reported by Rabi \cite{Rabi},
who studied the time dependent dynamics of a spin subject to an external
time-varying magnetic field. The phenomenon is presently considered as a
tool in characterizing coherence (and decoherence) of quantum states. Due to
the possibility offered by superconducting circuit integration, recent
attention has been devoted to detect Rabi oscillations in macroscopic
quantum systems based on Josephson junctions. The reported experiments \cite
{Martinis02,Claudon04} indicate evidence of phase oscillations in Josephson
junctions similar to what can be expected from the transitioning between the
quantum states of a multi-level atom. In this letter we demonstrate that the
energetic properties of different dynamical states in a nonlinear classical
Josephson junction can produce phase oscillations with observable features
analogous to the reported quantum phenomena.

The above mentioned investigations of Rabi oscillations \cite
{Martinis02,Claudon04} were concerned with excitations of the zero-voltage
of Josephson tunnel junctions driven by weak external microwave signals. The
properties of dc and ac driven Josephson junctions have been investigated in
several contexts \cite{Barone82,Vanduzer98}, but most of the attention has
been devoted to the phase-locking of nonzero dc-voltage states such as
constant-voltage Shapiro steps in the current-voltage characteristics. Our
investigation is entirely focused on the phase-locking of a Josephson
junction in the zero-voltage state. We assume that the relevant equation of
motion for this system is given by the classical RSCJ (Resistively and
Capacitively Shunted Junction) model \cite{Barone82,Vanduzer98}, 
\begin{eqnarray}
\ddot \varphi +\alpha \dot \varphi +\sin \varphi =\eta +\varepsilon _s\sin
\omega _st\;.
\end{eqnarray}
Here $\sin \varphi $, $\eta $, and $\varepsilon _s\sin \omega _st$ represent
Josephson, dc-bias, and ac-bias currents, respectively, normalized to the
Josephson critical current $I_c$. The amplitude and temperature dependence
of $I_c$ is established by the Ambegaokar-Baratoff equation \cite
{Barone82,Vanduzer98}. Time is normalized to the inverse of the Josephson
plasma frequency $\omega _0^{-1}$, where $\omega _0^2=2eI_c/\hbar C=2\pi
I_c/\Phi _0C$, $\Phi _0=h/2e=2.07\cdot 10^{-15}Wb$, and $%
C$ the capacitance of the junction.
Damping is represented through
$\alpha =\hbar \omega _0/2eRI_c=\Phi _0\omega _0/2\pi RI_c$,
where $R$ is the model resistance.
Normalized energy is defined by 
\begin{eqnarray}
H=\frac 12\dot \varphi ^2+1-\cos \varphi -\eta \varphi \;,
\end{eqnarray}
where the characteristic energy is $H_J=I_c\Phi _0/2\pi $, and the
resulting input power is
\begin{eqnarray}
\dot H=\varepsilon _s\dot \varphi \sin \omega _st-\alpha \dot \varphi ^2\;.
\end{eqnarray}

Our analysis of phase-locking in the system described above take into
account the anharmonicity of the potential through the zero-voltage,
phase-locked monochromatic ansatz, 
\begin{eqnarray}
\varphi =\varphi _0+\psi \;=\;\varphi _0+a\sin (\omega _st+\theta )\,,
\end{eqnarray}
where $\langle \dot \varphi \rangle =\dot \varphi _0=0$, and $a$ is an
oscillation amplitude. Inserting this ansatz into Eq.~(1) gives the
static, 
\begin{eqnarray}
J_0(a)\sin \varphi _0=\eta
\end{eqnarray}
component, where $J_n$ is the Bessel function of the first kind and $n$th
order. The dynamic component is
determined by 
\begin{eqnarray}
\ddot \psi +\alpha \dot \psi +\cos \varphi _0\sin \psi =\varepsilon _s\sin
\omega_st \; .
\end{eqnarray}
Inserting (4) and (5) into (6) gives the steady-state nonlinear
relationships, 
\begin{eqnarray}
\varepsilon _s^2 &=&a^2\left[ \left( \omega _s^2-\omega _r^2\right)
^2+\left( \alpha \omega _s\right) ^2\right] \\
\tan \theta _0 &=&\frac{\alpha \omega _s}{\omega _s^2-\omega _r^2} \; ,
\end{eqnarray}
where $\omega _r$ is the anharmonic resonance frequency
\begin{eqnarray}
\omega _r^2=\frac{2J_1(a)}a\sqrt{1-\left( \frac \eta {J_0(a)}\right) ^2} \; .
\end{eqnarray}
Notice that $\omega _r$ is the natural anharmonic frequency of an excitation
with amplitude $a$ when $\varepsilon _s=\alpha =0$. Thus, the solution (for $%
\varepsilon _s=\alpha =0$) establishes a useful ansatz, 
\begin{eqnarray}
\varphi =\varphi _0+a\sin (\omega _rt+\theta ) \; ,
\end{eqnarray}
for the unperturbed problem if $\varphi_0$ is given by Eq.~(5).

We substantiate this very simple ansatz through Fig.~1, which compares
direct numerical simulations of Eq.~(1) with the predicted
relationships (9) and (5), as well as the energy $H-H_0$,
where $H_0=1-\sqrt{1-\eta^2}-\eta\sin^{-1}\eta$, as a function of $a$.
The average total energy for the ansatz Eq.~(10) is 
\begin{eqnarray}
\bar{H} = \frac14a^2\omega_r^2+1-\omega_r^2\frac{aJ_0(a)}{2J_1(a)}%
-\eta\sin^{-1}\left(\frac{\eta}{J_0(a)}\right) \; .
\end{eqnarray}
Specific system parameters are: $\eta=0.94259$ and $\alpha=\varepsilon_s=0$.
We clearly observe that the ansatz represents the
anharmonicity very well for small to moderate amplitudes.

\begin{figure}[tbp]
\centering
\includegraphics[width=3.40in,angle=0]{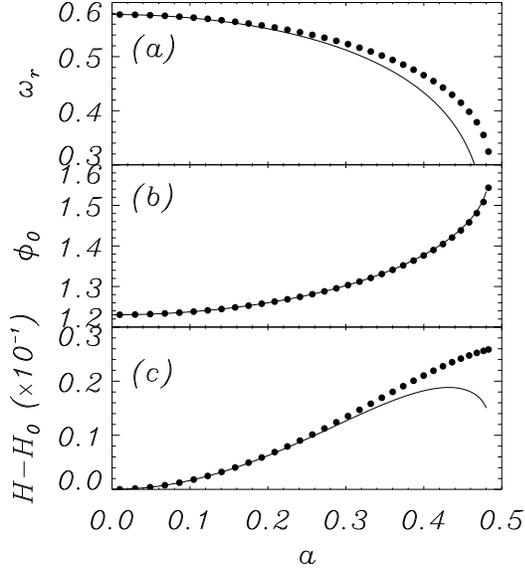}
\vspace{-0.5 in}
\caption{Comparison between numerical results ($\bullet$) and analytical
ansatz Eq.~(10) (solid lines) for anharmonic relationships between (a)
frequency and amplitude Eq.~(9), (b) average phase and amplitude Eq.~(5),
and (c) energy and amplitude Eq.~(11). Data shown for $\eta=0.94259$ and $%
\alpha=\varepsilon_s=0$.}
\label{fig:fig1}
\end{figure}

Figure 2 shows the comparisons between the perturbation analysis
and numerical simulations for the phase-locked
states when $\alpha=10^{-3}$, $\eta=0.94259$ for two different frequencies $%
\omega_s=\sqrt[4]{1-\eta^2}=0.577886$ (Fig.~2a,b) and $\omega_s=0.5237246$
(Fig.~2c,d), the latter being the resonance frequency for an amplitude close
to the switching point from the zero-voltage state \cite{Jensen_MQC2}. We
clearly see the close agreement between theory and simulations outlining the
anharmonic saturation of the resonance in Fig.~2a,b, and the switching
between small amplitude (off-resonance) and large amplitude (resonant)
excitations in Fig.~2c,d. The average energy for the phase-locked system is 
\begin{eqnarray}
\bar H=\frac14a^2\omega_s^2+1-\omega _r^2\frac{
aJ_0(a)}{2J_1(a)}-\eta\sin^{-1}\left( \frac \eta {J_0(a)}\right) \; .
\end{eqnarray}
Thus, the phase-locked states seem adequately described by the anharmonic
perturbation method outlined above. We will later use $\omega_s=0.577886$ as
the signal and $\omega_p=0.5237246$ as the probe frequency for demonstrating
the classical equivalent to Rabi oscillations.

\begin{figure}[tbp]
\centering
\vspace{-0.4 in}
\includegraphics[width=3.40in,angle=0]{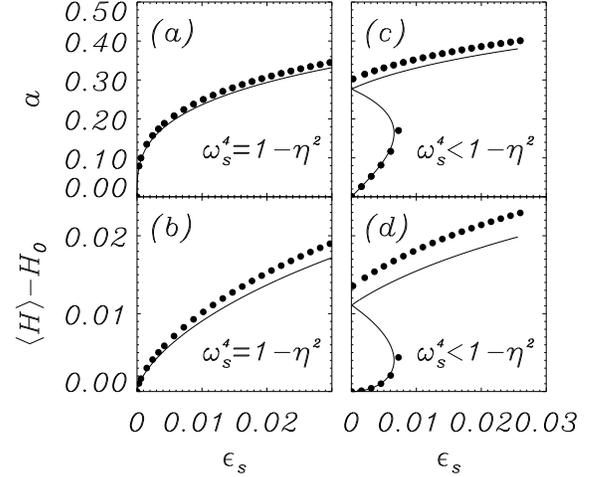}
\vspace{-0.7 in}
\caption{Oscillation amplitude $a$ and energy $\langle H\rangle-H_0$ as a
function of microwave amplitude $\varepsilon_s$ for $\alpha=10^{-3}$ and
signal frequencies (a-b) $\omega_s=\sqrt[4]{1-\eta^2}=0.577886$, and (c-d) $%
\omega_s=0.5237246$. Numerical simulations of Eq.~(1) are represented by
markers ($\bullet$), and solid lines represent Eqs.~(7) and (12).}
\label{fig:fig2}
\end{figure}

Stability of the phase-locked state can be evaluated by following the
treatment outlined for breathers in driven sine-Gordon systems \cite{mrsngj}%
. The energy change during one microwave period is 
\begin{eqnarray}
&&\Delta H  \; = \; \int_0^{\frac{2\pi }{\omega _s}}\dot Hdt \; = \; \\
&&-\varepsilon _sa\pi \sin \theta -\alpha \pi a^2\omega _s \; = \;
I^{(in)}\sin \theta -I^{(out)} \; ,
\end{eqnarray}
where Eq.~(14) is the result for the monochromatic ansatz, and where steady
state phase-locking is characterized by, 
\begin{eqnarray}
\Delta H &=&0\;\;\Rightarrow \; \;
\sin \theta _0 \; = \;
\frac{I^{(out)}}{I^{(in)}}\;=\;-\frac{\alpha a\omega _s}{\varepsilon _s} \; .
\end{eqnarray}
Notice that this power-balance expression is consistent with
(7) and (8).

A perturbation to a phase-locked state may introduce slow (transient or
steady state) modulation frequencies to the dynamics. Such modulations have
also been observed and described in the context of breather stabilization in
perturbed sine-Gordon equations \cite{mrsngj}. We give specific results
below for the phase-locked RSCJ model.

We assume that a modulation frequency to the phase-locked state is slow and
that the amplitude is small, i.e., that the modulated phase is given by 
$\theta =\theta _0+\delta \theta$,
with $|\delta \theta |\ll 1$ and $|\dot {\delta \theta }|\ll \omega _s$. In
this approximation, we can write the change in total energy over one driving
period as 
\begin{eqnarray}
&&\Delta H \; = \;
\frac{2\pi }{\omega _s}\dot{\bar H}\;=\;\frac{2\pi }{\omega _s}
\frac{\partial \bar H}{\partial a}\frac{\partial a}{\partial \omega _r}%
\,\ddot {\delta \theta } \; = \; I^{(in)}\cos \theta _0\,\delta \theta 
\nonumber \\
&&\Rightarrow \;\;\frac 2{\omega _s}\frac{\partial \bar H}{\partial a}\frac{%
\partial a}{\partial \omega _r}\,\ddot {\delta \theta } \; =\; a\sqrt{%
\varepsilon _s^2-(a\alpha \omega _s)^2}\;\delta \theta \; .
\end{eqnarray}
Equation (16) provides a slow modulation frequency $\Omega_R$ given by 
\begin{eqnarray}
\Omega _R^2=-\frac{1}{2}\omega_sa\sqrt{\varepsilon_s^2-(a\alpha\omega_s)^2}\;%
\frac{\partial \omega _r}{\partial a}\left( \frac{\partial \bar H}{\partial a%
}\right) ^{-1}
\end{eqnarray}
where $a$ and $\varepsilon_s$ are related through Eq.~(7), and $a$ is
related to $\omega_r$ through Eq.~(9). Notice that $\partial\omega_r/%
\partial a < 0$, leaving Eq.~(17) with a (real) positive value.

\begin{figure}[tbp]
\centering
\vspace{-0.2 in}
\includegraphics[width=3.40in,angle=0]{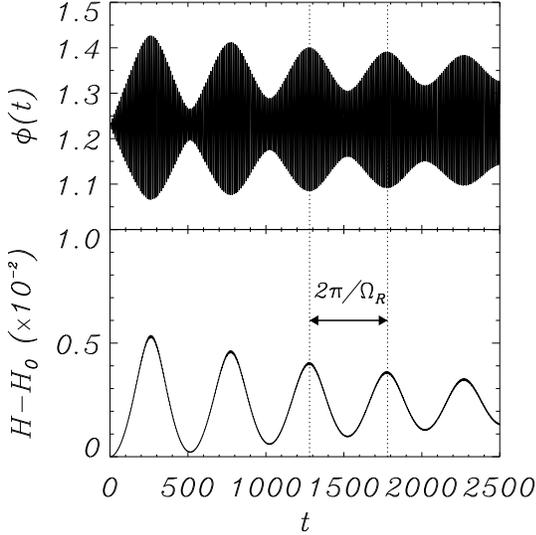}
\vspace{-0.7 in}
\caption{Frequency/amplitude and energy modulation as a function of time
after the onset of a microwave field with amplitude $\varepsilon_s=10^{-3}$.
Other system parameters are $\alpha=10^{-3}$, $\omega_s=\sqrt[4]{1-\eta^2}$,
and $\eta=0.94259$. Results are obtained from Eq.~(18).}
\label{fig:fig3}
\end{figure}
\begin{figure}[tbp]
\centering
\vspace{-0.2 in}
\includegraphics[width=3.40in,angle=0]{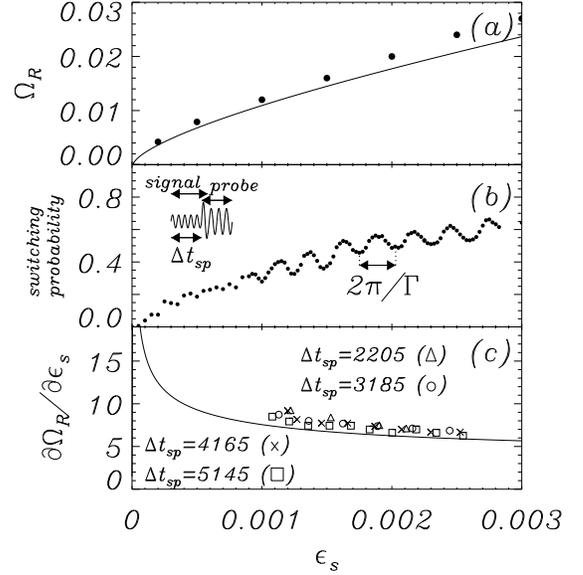}
\vspace{-0.5 in}
\caption{Dynamic Modulation as a function of microwave amplitude
$\varepsilon_s$.
Parameters are: $\eta=0.94259$, $\alpha=10^{-3}$, $\omega_s=\sqrt[4]{1-\eta^2%
}=0.577886$. (a) Modulation frequency $\Omega_R$.
Numerical simulations of Eq.~(18) are
represented by markers ($\bullet$) and the solid line represents Eq.~(17).
(b) Switching probability. Each point represents 10$^4$ events simulated through
Eq.~(20) with parameters: $\omega_p=0.5237246$, $\varepsilon_p=0.004$, $%
k_BT/E_J=5\cdot10^{-5}$, $t_s^\uparrow=10^4$, $t_s^\downarrow=t_s^%
\uparrow+4043$, $t_p^\uparrow=t_s^\uparrow+3185$, $t_p^\downarrow=t_s^%
\uparrow+6248$. All switching events take place in the interval $%
]t_p^\uparrow;t_p^\downarrow]$. (c) Variation in switching
probability. Solid line
represents the variation as derived from Eq.~(17). Markers represent
simulations of Eq.~(20), analysed through Eq.~(21), with parameters and
microwave timings $t_s^\downarrow-t_p^\uparrow$ and $t_p^\downarrow-t_p^%
\uparrow$ set as in (b), but with probe onset times given by $\Delta
t_{sp}=2205$ ($\triangle$), $\Delta t_{sp}=3185$ ($\circ$), $\Delta
t_{sp}=4165$ ($\times$), and $\Delta t_{sp}=5145$ ($\Box$).}
\label{fig:fig4}
\end{figure}

This modulation can be directly observed as transient behavior for $t>0$ in
the equation 
\begin{eqnarray}
\ddot \varphi +\alpha \dot \varphi +\sin \varphi =\eta +\Theta
(t)\varepsilon _s\sin (\omega _st+\theta _s)\;.
\end{eqnarray}
where $\Theta(t)$ is Heaviside's step function, and $\theta_s$ is a
constant, but random, phase. Figure 3 shows an example for $\alpha=10^{-3}$, 
$\eta=0.94259$, $\omega_s=0.577886$, and $\varepsilon_s=10^{-3}$. We observe
clear evidence of the slow modulation frequency $\Omega_R$ around the
phase-locked state at $\omega_s$ of phase, amplitude, and energy. This
behavior is present regardless of the value of $\theta_s$. Thus, for any
value of $\theta_s$ we can expect that the energy is at its minimum $H_0$
for $t\le0$, and oscillates around $\bar{H}$ for $t>0$ with frequency $%
\Omega_R$. Combining this information, we can now write the time evolution
of the energy as 
\begin{eqnarray}
H(t)=\left\{ 
\begin{array}{ccc}
H_0 & , & t\le 0 \\ 
\bar H-(\bar H-H_0)e^{-\beta t}\cos \Omega _Rt & , & t>0
\end{array}
\right.
\end{eqnarray}
where $\beta^{-1}$ is the transient time \cite{transient_time}.

Figure 4a shows a direct comparison between numerical simulations
and the perturbation result Eq.~(17) of the modulation
frequency as a function of the microwave amplitude $\varepsilon_s$. The
overall agreement is very good, and the discrepancy is
understandable considering the monochromatic ansatz, which neglects energy
contributions from higher harmonics.

The above analysis can be applied directly to recent
experimental reports \cite{Martinis02,Claudon04} of Rabi oscillations. A
complete system with signal and probe is governed by
\begin{eqnarray}
&&\ddot \varphi +\alpha \dot \varphi +\sin \varphi =\eta +(\Theta(t_s^{\uparrow })-\Theta (t_s^{\downarrow }))\varepsilon _s\sin (\omega_st+\theta _s) \nonumber \\
&& + (\Theta (t_p^{\uparrow })-\Theta (t_p^{\downarrow}))\varepsilon _p\sin (\omega _pt+\theta _p)+n(t)\;,
\end{eqnarray}
where $t_s^{\uparrow }<t_s^{\downarrow }$, $t_s^{\uparrow }<t_p^{\uparrow
}<t_p^{\downarrow }$, and $n(t)$ is a thermal noise term determined by the
fluctuation-dissipation relationship \cite{Parisi_88}. Following the
experimental procedure, we choose the probe frequency smaller than the
signal frequency, $\omega _p<\omega _s\approx \sqrt[4]{1-\eta ^2}$, and $%
\varepsilon _p$ is chosen such that phase-locking at $\omega _p$ presents
two possible energy states (see Fig.~2c,d), one small amplitude and one near
or beyond switching. Switching is then induced with high probability if the
probe field is causing a high energy state at $\omega _p$, while low
probability of switching occurs if the probe field excites the low energy
state (see Fig.~2d). The determining factor is the timing $\Delta
t_{sp}=t_p^{\uparrow }-t_s^{\uparrow }$ between the onset of the two
microwave fields. This is apparent from Fig.~3 and Eq.~(21) since $\Omega
_R\Delta {t_{sp}}=2\pi p$ ($p$ being an integer) will provide for a
relatively low onset energy for the probe field with subsequent low
probability for exciting a high-energy state at $\omega _p$. However, for $%
\Omega _R\Delta {t_{sp}}=2\pi p+\pi $, the onset energy of the system is
relatively large, and a much higher probability for high energy excitation
at $\omega _p$ can be expected. A reasonable set of probe parameters can be
chosen to $\omega _p=0.5237246$ and $\varepsilon _p=0.004$. As mentioned
above, this probe frequency provides a resonant amplitude that takes the
junction phase close to the saddle point of the energy potential, and the
amplitude $\varepsilon _p$ is chosen such that both high and low energy $%
\omega _p$-states can be chosen based on the state of the system at time $%
t=t_p^{\uparrow }$. This is the recipe for the classical equivalent of Rabi
oscillations in microwave driven Josephson junctions, and they can be
observed in several different ways. One is to keep $\varepsilon _s$ constant
while varying $\Delta t_{sp}$ for the direct observation of the modulation
frequency as has been done in \cite{Claudon04}. Another is to keep $\Delta
t_{sp}$ constant while varying $\varepsilon _s$, as was demonstrated in \cite
{Martinis02}. We will demonstrate the latter in Fig.~4b, where we show
simulation results for $\Delta t_{sp}\approx 3185$. Each point on the figure
represents 10,000 events of sequentially applying the signal and the probe,
each with random phases $\theta _s$ and $\theta _p$, and detecting whether
or not the system switched to the non-zero voltage state within the probe
duration $t_p^{\downarrow }-t_p^{\uparrow }$. The fraction of
switching events is shown in Fig.~4b. We note that this set of data has
been generated with a small amount of thermal noise ($k_BT/E_J\approx 5\cdot
10^{-5}$) in order to mimic experimental reality. However, this is not
significantly affecting the result, since the random phases $\theta _s$ and $%
\theta _p$ are the primary factors in determining if a switching event
occurs. The variations in switching probability as a function of $%
\varepsilon _s$ are very clear from this figure, and the characteristics of
the observed variation is given by 
\begin{eqnarray}
\Gamma =\Delta t_{sp}\frac{\partial \Omega _R}{\partial \varepsilon _s}\;,
\end{eqnarray}
where $\Gamma $ is defined in Fig.~4b. A convenient way of quantitatively
comparing the modulation in the switching probability as a function of
microwave amplitude $\varepsilon _s$ to the predicted behavior related to
the modulation frequency $\Omega _R$ is to show $\Gamma /\Delta
t_{sp}=\partial \Omega _R/\partial \varepsilon _s$ as a function of $%
\varepsilon _s$ for different $\Delta t_{sp}$. This is done in Fig.~4c,
where the solid curve is derived from Eq.~(18), and the markers represent
simulation results as shown in Fig.~4b. Good consistent agreement is found between the
classical theory and the modulation of the probability of switching in the
range where we have obtained data. The trend in the comparison, i.e., all
the simulation results being above the theory line, is directly related to
the monochromatic ansatz, which underestimates the modulation frequency
(Fig.~4a) and the system energy (Fig.~2) of the phase-locked state. The
above analysis and simulations demonstrate a direct analogy between
transient modulation in a classical nonlinear Josephson junction irradiated
with microwaves and reported Rabi oscillations in a system under similar
conditions.

In conclusion, we have shown that energy level degeneration in the nonlinear
system describing a Josephson junction can lead to transient oscillations
when the phase of the junction is modulated by adequately shaped external
ac-current terms. The resulting dynamical features demonstrate close analogy
to the quantum phenomenon of Rabi oscillations. We note that a similar
analogy between classical and quantum analysis in relation to Rabi
oscillations and splitting has been investigated in quantum optics \cite
{zhu90}. As far as Josephson effect is concerned, analogies between
classical and quantum descriptions of experimental results have been
demonstrated by recent comparisons of theoretical and experimental results 
\cite{Jensen_MQC2,prlngj04} relating to resonant switching from zero-voltage
states.

This work was supported in part by the UC Davis Center for Digital Security
under AFOSR grant FA9550-04-1-0171. We are
grateful to Prof.~Alan Laub for carefully reading the manuscript.


\begin{thebibliography}{99}
\bibitem{transient_time}  The transient time $\beta^{-1}$ can be estimated
from the small perturbation behavior in Eq.~(6) at the frequency $\omega_s$.
\end{thebibliography}
\end{document}